\documentclass[12pt,preprint]{aastex}

\slugcomment{Accepted by Ap. J. in April 2004.}

\shorttitle{Giant Micropulses and Pulses from Pulsars}
\shortauthors{I.H. Cairns}

\authoraddr{Iver H. Cairns, 
School of Physics, University of  Sydney, NSW 2006, 
Australia. (e-mail: cairns@physics.usyd.edu.au)} 

\begin{document}


\title{Properties and Interpretations of Giant Micropulses and 
Giant Pulses from Pulsars}


\author{Iver H. Cairns}

\affil{School of Physics, University of  Sydney, Sydney,  
Australia.}


\begin{abstract}
Giant pulses and giant micropulses from pulsars are distinguished 
from normal pulsed emission by their large fluxes, rarity,  
approximately power-law distribution of fluxes and, 
typically, occurrence in restricted phase windows. Here existing 
observations of flux distributions are manipulated into  
a common format and interpreted in terms of theories 
for wave growth in inhomogeneous media, with the aim of constraining 
the emission mechanism and source physics for giant pulses and micropulses. 
Giant micropulses near 2 GHz (PSRs B8033-45 and B1706-44) and 
0.4 GHz (PSR B0950+08) have indices $\alpha = 6.5 \pm 0.7$ for the 
probability distribution $P(E)$ of the electric field $E$, with 
$P(E) \propto E^{-\alpha}$. Giant pulses (PSRs B0531+24, B1937+214, 
and B1821-24) have $\alpha$ ranging from $4.6 \pm 0.2$ to $9 \pm 2$, 
possibly increasing with frequency. 
These are similar enough to regard giant micropulses and 
pulses as a single phenomenon with a common physical explanation. 
The power-law functional form and values of $\alpha$ observed are 
consistent with predictions for nonlinear 
wave collapse, but inconsistent with known self-organized critical 
systems, nonlinear decay processes, and elementary burst theory. 
While relativistic beaming may be important, its statistics are yet  
to be predicted theoretically and collapse is currently the favored interpretation.  Other possibilities remain, including stochastic 
growth 
theory (consistent with normal pulse emission) and, less plausibly, refractive lensing.  Unresolved issues remain for all four  
interpretations and suggestions for further work are given. 
The differences between normal and giant pulse emission 
suggest they have 
distinct source regions and emission processes. 

%  existing collapse theory 
%must still be extended to pulsar plasma conditions and issues 
%resolved relating to magnetization and dimensional effects, the possible 
%variation of $\alpha$ with frequency, and radiation mechanisms 
%associated with collapsing wavepackets. Nevertheless, required future 
%observational work includes obtaining extended datasets with a larger 
%range of fluxes and better statistics, so as to confirm that the apparently 
%power-law distributions observed are intrinsic and not the
%result of vector convolution of two non-power-law wave populations 
%(as found in some phase ranges for normal emission from some pulsars). 
%This future work will also permit the relevance of stochastic growth theory 
%(SGT) to be properly assessed for giant phenomena; since SGT appears 
%consistent with normal pulse emission, the differences between normal and giant 
%emission suggest they have distinct source regions and that collapse, not SGT, 
%is most favored for understanding giant pulses and micropulses. 
\end{abstract}

%\keywords{pulsars: general --- pulsars: individual (Vela) --- methods: statistical --- plasmas --- radiation mechanisms: non-thermal --- stars: neutron}


%\keywords{methods: statistical --- plasmas --- pulsars: general --- 
%pulsars: individual (Crab, Vela, B0950+08, B1706-44, B1937+214, 
%B1821-24) --- radiation processes: non-thermal --- waves }


\section{Introduction}

  Giant pulses from the Crab pulsar (B0531+24), defined as pulses whose 
pulse-integrated fluxes $F_{tot}$
exceed ten times the average pulse-integrated flux $F_{av,tot}$, were intrinsic to 
discovery of the Crab pulsar 
\citep{staelinreifenstein1968,heilesetal1970} and 
have been studied extensively since 
\citep{lundgrenetal1995,sallmenetal1999,hankinsetal2003}. 
They are remarkable for several reasons other than their very large 
fluxes, which are sometimes hundreds to thousands of times 
larger than $F_{av,tot}$. First, only the Crab pulsar and the 
millisecond pulsars B1937+214 and B1821-24 
\citep{cognardetal1996,romanijohnston2001} are known to 
produce giant pulses. Second, giant pulses are produced sometimes 
in the same phase windows as the normal pulses, for the Crab pulsar 
\citep{staelinreifenstein1968,heilesetal1970} and PSR B1821-24 
\citep{romanijohnston2001}, but also in narrow phase windows 
trailing the main pulse and interpulse for PSR B1937+214 
\citep{cognardetal1996}. Third, their 
distribution of fluxes appears to be a power-law and is usually 
considered to be clearly distinguishable from that for normal 
pulses \citep{lundgrenetal1995,cognardetal1996}. Indeed, the 
distribution of pulse-integrated 
fluxes for normal pulses often appears approximately Gaussian 
when binned in the logarithim of the flux [e.g.,  
\citep{johnstonetal2001}], and so is approximately lognormal, 
and phase-resolved flux distributions for normal pulses are 
often interpretable in terms of a lognormal distribution or the 
vector convolution of a lognormal with either a Gaussian or lognormal 
distribution \citep{cetal2001,cetal2002a,cetal2002b,cetal2002c}. 
The importance 
% cetal2002c
of the (electric) field statistics is that they are determined by the 
source physics, emission mechanism, and propagation effects 
\citep{cetal2001,cetal2002a}, as well as relativistic beaming 
effects: accordingly, different flux distributions imply different 
emission mechanisms and/or 
source regions for giant pulses and the normal pulses, as surmised 
on intuitive grounds previously \citep{romanijohnston2001}. 

   Giant micropulses have been discovered very recently for the Vela 
pulsar (B8033-45) and pulsars B1706-44 and B0950+08 
\citep{johnstonetal2001,krameretal2002,johnstonromani2002,cetal2002c}. 
These are short-duration intense events that have phase-resolved 
fluxes much greater than ten times the average flux for that phase, 
but have pulse-integrated fluxes less than 10 times the average and 
so are not giant pulses. Importantly, giant micropulses also 
appear to have approximately power-law 
distributions of flux, are rare, do not occur for every pulsar \citep{johnstonromani2002}, 
and occur in narrow phase windows either leading [Vela \citep{johnstonetal2001} and B0950+08 \citep{cetal2002c}] 
or trailing [PSR B1706-44 \citep{johnstonromani2002}] the average 
pulse peak.  These features 
all point to strong similarities between giant micropulses and giant 
pulses. Indeed, it is sometimes suggested that giant 
pulses and micropulses are generated near the light cylinder, in a similar 
fashion  to the outer-gap model for X-ray and/or gamma ray pulses 
\citep{romaniandyadigaroglu1995,romanijohnston2001,johnstonromani2002}.
This paper compares the observed 
field statistics with theoretical predictions to constrain the 
emission mechanism responsible for giant pulses and giant micropulses.      

   Growth of plasma waves and radiation in inhomogeneous plasmas, and its 
propagation between source and observer,  naturally results in bursty, 
time-variable radiation. Multiple theories exist for the 
associated field statistics, differing due to 
varying degrees of self-consistent interaction between the waves, driving 
particles, and background plasma, the detailed emission mechanisms 
producing the waves, the importance of scattering,
and the number of wave populations contributing. The giant and ``normal'' components 
of pulsar radio emissions have been discussed in 
terms of several collective (or coherent) emission mechanisms, including: (i) linear 
plasma instabilities such as curvature 
emission and beam instabilities [e.g., \citet{melrose1996,gedalinetal2002}], 
(ii) linear mode conversion \citep{melrosegedalin1999,cetal2001}, 
and (iii) nonlinear modulational 
instabilities and collapse of wavepackets (Asseo, Sol, \& Pelletier 
1990, Asseo 1996, Weatherall 1997, 1998, Hankins et al. 2003). 
Theoretical frameworks considered include self-organized criticality [SOC] 
\citep{baketal1987,bak1996,youngkenny1996}, stochastic growth theory [SGT] 
\citep{r1992,cr1999,rc2001,cetal2001,cetal2002a,cetal2002b}, 
and nonlinear structures (Pelletier, Sol, \& Asseo 1988, Asseo 1996, Weatherall 1997, 1998). 
Other possibilities include refractive lensing events due to propagation 
through density inhomogeneities (M.A. Walker, personal communication, 2003) 
and time-variable relativistic beaming effects.  
As summarized below, these emission mechanisms and theoretical frameworks 
predict different field statistics, so that comparisons between theory and 
observation permit the emission mechanism and source physics to be 
constrained. Recent analyses do this for normal pulses of the Vela pulsar 
%\citep{cetal2001,cetal2002a,cetal2002b}
and pulsars B1641-45 and B0950+08 
\citep{cetal2001,cetal2002a,cetal2002b,cetal2002c}. 

   The goals of this paper are to: (1) show that the distributions 
of electric field strengths for known sources of giant pulses and 
giant micropulses are all 
approximately power-law at high fluxes and have sufficiently 
similar power-law indices for them to be regarded as one population, 
(2) constrain which models for emission processes and 
source physics for giant pulses/micropulses remain viable, by 
comparing observations with theoretical predictions, (3) show that 
wave collapse is the most favored interpretation for giant pulses and micropulses, in the absence of detailed calculations for 
relativistic beaming, while 
some other interpretations appear viable but less plausible and 
others are inconsistent with available data, and (4) 
point out limitations in current theories for wave collapse. These goals are 
addressed by summarizing theory for field 
statistics (Section 2), analyzing and discussing the field statistics of giant 
micropulses (Section 3) and giant pulses (Section 4), 
and comparing the observations with theoretical predictions (Section 5). 
The results are summarized and brief conclusions given in Section 6. 

\section{Summary of Relevant Theories for Field Statistics}
 
   The pulsar plasma is expected to be significantly time-variable. 
Accordingly, relativistic beaming effects might lead to 
significant variations 
in the observed fields due to changes in the Lorentz factor $\Gamma$ 
and the direction of primary emission relative to a stationary 
observer, thereby potentially modifying the intrinsic field 
statistics of a specific mechanism in the source rest frame. 
Detailed reasons include the following. First, the observed flux 
varies as $\Gamma^{-3}$ \citep{weatherall1997}, where 
$\Gamma$ is the Lorentz factor. Second, the angular 
distribution of radiation becomes increasingly directed in the 
forward direction (along ${\bf v}$) as $\Gamma$ increases. 
Third, temporal changes in the emission direction relative to a stationary 
observer may occur: for instance, Alfven waves or other disturbances 
might cause the bundle of magnetic field lines carrying an active 
emission source to move the relativistic beaming pattern into an 
observer's viewing window at some times and not at other times. 
Unfortunately predictions do not yet exist for the distributions $P(F)$ resulting from variations in $\Gamma$ and the 
relative viewing angle, which will depend at least on the statistics of 
$\Gamma$, the source location, and the variability of the magnetic 
field and flow velocity of the source plasma. 
 
%Accordingly, changes in $\Gamma$ and the direction of maximum 
%emission relative to a stationary observer might modify the intrinsic 
%field statistics.  
  A qualitative argument suggests that relativistic beaming 
effects for a single emission mechanism and single source 
plasma can not determine the field statistics of all 
components of pulsar emission: if they 
did, then the field statistics of all emission components would 
have the same functional form. This is contrary to normal 
pulsar emission having lognormal statistics 
\citep{cetal2001,cetal2002a,cetal2002b,cetal2002c} 
and giant pulses and micropulses having power-law statistics, 
as described below. Instead, if all components involve relativistic 
beaming then it seems clear that these 
different statistics must require different source conditions (e.g., distributions of $\Gamma$ and/or relative emission angle) and/or 
different emission mechanisms. It is undoubtedly true that 
relativistic beaming effects are important for pulsar emissions. Nevertheless, in the current absence of theoretical 
calculations that show how relativistic beaming can modify 
intrinsic field statistics or produce qualitatively 
different field statistics for different source conditions 
or mechanisms, attention is focused below on 
mechanisms that do not involve relativistic beaming effects.

    Several theories exist for wave growth in inhomogeneous media, 
each predicting different field statistics which can 
be used to constrain the source physics and emission mechanism, 
including whether nonlinear processes are active. These theories, 
their justifications, and predictions are detailed elsewhere 
(Bak et al., 1987; Robinson \& Newman 1990; Robinson 1992, 1995, 1997; 
Robinson et al. 1993; Bak 1996; Cairns \& Robinson 1999; Robinson 
\& Cairns 2001; Cairns et al., 2001, 2003a) 
and only a brief summary is given here.  
Define the probability distribution $P(E)$ 
of the (electric) field $E$ 
and the related distributions $P(E^{2}) \propto P(F)$ of the 
field energy $E^{2}$ and flux $F$, with $F \propto E^{2}$. These distributions 
are normalized according to $\int P(X) dX = 1$ where $X = E$, $E^{2}$, or $F$. 

     SOC involves fully self-consistent interactions near marginal stability between 
the waves, driving particles, and the background plasma \citep{baketal1987,bak1996}: 
the system is driven away from stability but then relaxes 
towards marginal stability, with no preferred scales over large ranges of 
distance or time. SOC predicts a power-law distribution of energy
releases over many decades with index $\beta \approx 1$, with a range 
$\approx 0.5 - 2$ in known systems \citep{bak1996}. 
Then, assuming that the energy releases are adequately 
represented by the individual flux samples measured over some 
integration time (implying that event 
duration is unimportant, an assumption that remains to be checked 
observationally), so that 
\begin{equation}
P(F) \propto F^{-\beta}\ ,
\label{flux_equation} 
\end{equation}
SOC predicts a power-law with  
\begin{equation}
P(E) \propto E^{-\alpha} \ . 
\label{power-law}
\end{equation} 
Since $2E P(E^{2}) = P(E)$ 
via the normalization conditions, (\ref{flux_equation}) and (\ref{power-law}) imply 
$\alpha = 2\beta - 1$, whence $\alpha \approx 1$ with a range $\approx 0 - 3$ in known 
SOC systems.  SOC has been suggested for giant 
pulses \citep{youngkenny1996} and also for X-ray variability of active galactic 
nuclei \citep{baketal1988}. 

   The nonlinear self-focusing process of modulational instability typically leads to 
wave collapse, in which a wavepacket collapses to smaller spatial scales while 
intensifying, as reviewed elsewhere \citep{r1997,weatherall1997}. 
Non-relativistic electron-proton simulations of Langmuir waves 
show that modulational instability leads to wave collapse in two or three 
dimensions with power-law statistics at high $E$ described by (\ref{power-law}) 
\citep{r1997}. These simulations and associated scaling theory 
show that the index $\alpha$ depends strongly on the dimensionality $D$ and 
shape of the collapsing wavepackets (isotropic versus prolate versus oblate 
shapes relative to the magnetic field direction
and whether the field energy $W \equiv E^{2}$ is above or below the mean 
energy $\langle W \rangle$, as summarized in Table \ref{table_theory} 
\citep{rn1990,r1996}. Simulations that include electromagnetic effects 
for strongly magnetized, relativistic, 
electron-positron plasmas appropriate to pulsars also show wave collapse 
proceeding (Weatherall 1997, 1998). However, although these collapse events 
are qualitatively very similar to those for the electron-proton simulations, 
the statistics of the collapsing fields have not been published. It might be 
objected that modulational instability can lead to stable solitons and 
not collapse. However, this appears to be true only 
under restrictive conditions: for instance, only in one dimension (rather 
than two or three) does modulational instability leads to stable solitons 
rather than collapse for the electron-proton simulations described above 
\citep{r1997}. Similarly, for the  
electron-positron plasmas above, purely electrostatic calculations suggest that 
solitons are modulationally stable \citep{pelletieretal1988,asseoetal1990} 
but inclusion of electromagnetic effects leads to wave collapse in 
simulations (Weatherall 1997, 1998). Accordingly, it is presumed 
that modulational instablity leads to wave collapse and 
power-law statistics with indices similar to those given in Table 
\ref{table_theory} for electron-proton calculations. Possible 
differences in the indices for pulsar magnetospheres are discussed 
in Section 5.2. 

%\clearpage

\begin{table}[t]
\caption{Power-law indices $\alpha$ for the $P(E) \propto E^{-\alpha}$ 
distribution obtained from numerical simulations and scaling theory 
for wave collapse (Robinson and Newman 1990, Robinson 1997). 
The dimension $D$ is either $2$ or $3$. }
\label{table_theory}
\begin{tabular}{cccc}
\hline \\
Range of $E$&Isotropic&Prolate&Oblate \\
\hline \\
$E < \langle W \rangle^{1/2}$&$-(2D - 1)$&$-(2D - 3)$&$-1$\\
$E > \langle W \rangle^{1/2}$&$D + 2$&$D + 3$&$2D + 1$\\
\hline  \\
\end{tabular}
\end{table}

%\clearpage

Table \ref{table_theory} shows that collapse should produce negative 
$\alpha$ (corresponding to $P(E)$ increasing with $E$) for field 
energies less than the mean value $\langle W \rangle$, whereas 
$\alpha$ should be positive above the peak in $P(E)$ with integer 
values in the range $4 - 7$ that depend significantly on $D$ and 
the wavepacket shape. Importantly, 
%unless $\alpha \approx 3$ corresponding to $D = 1$ and isotropic collapse, 
these indices for high $E$ are significantly larger than those for 
known SOC systems, so that systems involving collapse should usually 
be distinguishable from SOC systems. 
%For future reference it is noted that these collapse simulations  
%are for electron-proton, non-relativistic systems with Langmuir, 
%lower hybrid and upper hybrid waves which have isotropic, prolate, and oblate 
%shapes, respectively. The latter two modes are relevant when the magnetic field 
%strength is appreciable in the plasma. 
The origin of these collapsing 
wavepackets is usually assumed to be a plasma instability, thereby potentially 
involving SGT, as discussed next. 

   SGT \citep{r1992,retal1993,r1995,cr1999,rc2001} 
treats systems in which an unstable particle distribution and associated waves 
driven by an instability couple self-consistently in an 
independent, spatially-inhomogeneous medium, causing the wave-particle system 
to fluctuate stochastically about marginal stability. % The medium and the specific 
%wave-particle interactions determine the relevant distance and time scales. In 
%SGT the wave gain $G \propto \log E$ is a stochastic variable and so, via the 
%Central Limit Theorem, pure SGT predicts lognormal statistics for $E$. That is, 
Pure SGT then predicts lognormal statistics for $E$:  
\begin{equation}
P(\log E) = (\sqrt{2\pi}\ \sigma)^{-1} e^{-(\log E - \mu)^{2} / 2\sigma^{2}}\ , 
\label{p_sgt}
\end{equation}
where $\mu = \langle \log E \rangle$ and $\sigma$ are the average and standard 
deviation of $\log E$, respectively, and $\log \equiv \log_{10}$ for future 
convenience. SGT is widely applicable in solar system plasmas 
\citep{retal1993,cr1999,cg2001,cm2001} and also describes very well the 
phase-resolved statistics of normal pulses from several pulsars 
\citep{cetal2001,cetal2002a,cetal2002b,cetal2002c}. 
 
   SGT can coexist with a nonlinear process active at high $E \ge E_{c}$. 
In this case, the $P(\log E)$ distribution is modified in two 
characteristic ways from the lognormal predicted for pure SGT : (i) if a decay 
process removes energy from the primary (parent) 
waves, then the $P(\log E)$ distribution is reduced near and above $E_{c}$ with 
known form \citep{retal1993,r1995,cg2001,cm2001}; (ii) if a nonlinear self-focusing 
process like wave collapse or modulational instability \citep{r1997} is active, increasing 
the number of intense wavepackets, then the field distribution is enhanced at high 
$E \ge E_{c}$ into a power-law tail with index given in Table \ref{table_theory}. 

  Thermal waves driven by an instability and subject to limited stochastic growth 
effects (but which have not evolved into a pure SGT state) have approximately 
power-law statistics at fields much greater than 
the average thermal level $E_{T}$ \citep{r1995,cetal2000}: the 
index $\alpha$ in (\ref{power-law}) is positive,  
depends on the difference between the average growth and damping rates 
divided by a stochastic parameter, and can take a wide range of possible 
values. Importantly, though, the majority of the fields will be within a 
few decades of $E_{T}$. 

Gaussian intensity statistics result from superposition of 
multiple random signals, as expected for 
measurement noise, sky background, and multiple unresolved subsources: 
\begin{equation}
P(I) = (\sqrt{2\pi}\ \sigma_{I})^{-1} e^{-(I - I_{0})^{2} / 2\sigma_{I}^{2}}\ .  
\label{p_i} 
\end{equation} 
In addition, closely Gaussian intensity statistics can result from scattering of 
radiation by density inhomogeneities between the source and observer 
under some circumstances \citep{ratcliffe1956,salpeter1967,rickett1977}. Refractive focusing 
can also lead to ``lensing'' events associated with caustics and 
other singularities (M.A. Walker, personal communication, 2003). 
In particular, Walker pointed out that the magnification depends 
on the geometric properties of caustics: he used established results 
from gravitational lensing theory [Equation (11.64a) 
of Schneider et al. (1992)] to show that the distribution of flux magnification factors $\mu = F/F_{0}$ for the lowest order 
critical curve (a fold) is power-law at large $\mu$, with 
\begin{equation}
P(\mu) \propto \mu^{-3}\ , 
\end{equation}
where $F_{0}$ is the original flux. Convolving this distribution with an 
initial distribution of fluxes leads to the result (\ref{flux_equation}) 
with $\alpha = 3$ at large $F$. Independently, the Referee pointed out 
that a direct analysis of lensing effects in scattering theory [Equation (28a) of \citet{salpeter1967}] yields the form (\ref{flux_equation}) 
with $\alpha = 3$. 

Predictions exist for 
the statistics of ``elementary burst'' systems 
\citep{rsw1996} and uniform secular growth \citep{cr1999}. However, these 
are considered very unlikely to be relevant and are not detailed here.  
   
\section{Statistics for Giant Micropulses} 

\subsection{PSR B1706-44}

  \citet{johnstonromani2002} observed giant micropulses from PSR B1706-44 at 1.5 GHz. 
They claimed that giant micropulses are separable from the normal pulse 
emission since at large $F$ the (non-standard) cumulative probability distribution 
CDF$(F)$, calculated using logarithmic binning in $F$ according to 
their Figure 4, changes to a power-law function with 
\begin{equation}
{\rm CDF}(F) = \int_{F}^{\infty} d(\log F) P(\log F) \propto F^{-\beta}  
\label{cdf_vela}
\end{equation}
and $\beta \approx 2.7 \pm 0.3$, the error being estimated by eye from their 
figure 4. [The usual cumulative probability distribution equals $1 - $CDF$(F)$ 
and $\log = \log_{10}$ here and below.] 
That figure shows the transition to an approximately power-law distribution 
occurring near $F = 10^{2.5}$ mJy, where $1$ jansky (Jy) equals 
$10^{-26}$ Wm$^{-2}$Hz$^{-1}$. 
 
   Consider now the   
field and intensity proxies $E'$ and $I'$, defined in terms of $F$ by  
\begin{eqnarray}
E' & = & \left( F / 1{\rm mJy} \right)^{1/2} \ , \\
I' & = &  F / 1 {\rm mJy}  \ . 
\end{eqnarray}
These proxies are proportional to the 
actual field $E$ and intensity $I$ incident on the antenna, which 
are used in (\ref{p_i}) and (\ref{p_sgt}), respectively,  and are used 
henceforth without prime symbols. 

Now, since $F \propto E^{2}$ 
and $P(\log E) = E P(E)$, it is easy to show that (\ref{cdf_vela}) leads to 
the form (\ref{power-law}) with $\alpha = 2\beta + 1 = 6.4 \pm 0.6$ for B1706-44's 
giant micropulses (Table \ref{table_pulses}). It is important to ascertain whether 
the flux observations are binned in linear rather than logarithmic bins: 
since linear binning would correspond to (\ref{cdf_vela}) in the 
form $\int dX P(X)$ with $X = F$ rather than $\log F$, and since 
$P(\log F) = F P(F)$, mistaking linear for logarithmic binning (or vice versa) leads 
to a difference of $2$ in the index $\alpha$. 

%\clearpage

\begin{table}[t]
\caption{Power-law indices $\alpha$ observed for the $P(E) \propto E^{-\alpha}$ 
distribution of giant pulses (GP) and giant micropulses (GM). Also given are the 
observing frequency and the observed range of the power-law statistics, measured 
in $\log_{10}$ of the flux $F$ in mJy (units of $10^{-29}$ Wm$^{-2}$Hz$^{-1}$). }
\label{table_pulses}
\begin{tabular}{ccccccc}
\hline \\
Pulsar&B1706-44&Vela&B0950+08&Crab&B1937+214&B1821-24 \\ \hline
Type&GM&GM&GM&GP&GP&GP \\
Frequency (GHz)&1.5&2.3&0.4&0.8&0.4&1.5 \\
Range of $\log F$&0.5&0.5&0.8&1.0&2.0&0.1 \\
$\alpha$&$6.4\pm 0.6$&$6.7\pm 0.6$&$6.2\pm 0.5$&$5.6\pm0.6$&$4.6\pm 0.2$&$9 \pm 2$ \\
\hline \\
\end{tabular}
\end{table}

%\clearpage

   Table \ref{table_pulses} also lists the range of $\log F$ 
for which power-law statistics are observed ($0.5$ for B1706-44), which might 
appear large enough for 
reasonable confidence to be attached to the inferred value of $\alpha$. However, 
as discussed in Section 5, apparently 
power-law statistics can arise due to vector superposition of lognormal and Gaussian 
distributions \citep{cetal2002d,cetal2002b}. 

\subsection{The Vela pulsar}

  Giant micropulses from the Vela pulsar at 2.3 GHz also have apparently power-law 
statistics at large $F$ that 
obey the form (\ref{cdf_vela}), this time with $\beta \approx 2.85 \pm 0.3$  
\citep{krameretal2002}, where the error is again estimated by eye.  Converting into 
the form (\ref{power-law}), as for 
B1706-44, leads to $\alpha = 6.7 \pm 0.6$. 

\subsection{Pulsar B0950+08}

   Giant micropulses from B0950+08 at 0.43 GHz have apparently 
power-law statistics at large $F$ that obey the form 
(\ref{power-law}) with index $\alpha = 6.2 \pm 0.5$ \citep{cetal2002c}. 
These giant micropulses are observed over a relatively large range 
of pulsar phase compared with those for Vela and B1706-44.  
Several other classes of emissions were identified with 
approximately power-law statistics. However, their indices 
were substantially less, with values in the range $1 - 4$. 

   Summarizing, the field distributions observed near 2 GHz and 
$0.4$ GHz for giant micropulses from three pulsars (Vela, B1706-44, 
and B0950+08) can be interpreted in terms of power-laws with high 
indices $\approx 6.5 \pm 0.7$. These limited data provide no 
evidence for $\alpha$ varying with observing frequency. 

\section{Field Statistics For Giant Pulses}

    Giant pulses from the Crab pulsar also have apparently power-law statistics at 
large $F$: binning linearly in $F$, the distribution is well described by 
\citep{lundgrenetal1995} 
\begin{equation}
P(F) \propto F^{-\gamma} 
\label{p_f_crab}
\end{equation}
with $\gamma = 3.3 \pm 0.3$ at $800$ MHz. Here $3.3$ is the most common 
value observed for the 
exponent and the ``error'' is the difference between the largest and smallest 
values observed \citep{lundgrenetal1995}. Then, since $P(E) = 2 E P(E^{2})$, 
(\ref{p_f_crab}) can be placed in the form (\ref{power-law}) with 
$\alpha = 2 \gamma - 1 = 5.6 \pm 0.6$. 

%    The power-law index for giant pulses from the Crab pulsar thus overlaps, within 
%both error bounds, those found for giant micropulses in Section 3. This suggests that 
%giant micropulses and giant pulses may share a common physical origin. 
 
   Giant pulses from the millisecond pulsar B1937+214 at 430 MHz 
also appear to be power-law distributed \citep{cognardetal1996}. Inspection of their 
Figure 3 suggests that the cumulative distribution is binned logarithmically, 
not linearly, whence (\ref{cdf_vela}) applies with $\beta = 1.8 \pm 0.1$ at $430$ MHz. 
Accordingly (\ref{power-law}) holds with $\alpha = 4.6 \pm 0.2$. [Note 
that linear binning would imply $\alpha = 2.6 \pm 0.2$, a clear 
difference.] 

   Taken together the results for the Crab and B1937+214 are consistent with 
giant pulses having power-law statistics with high indices $\approx 4.4 - 6.2$, 
and are only marginally inconsistent with them having a common index. Both possibilities, 
that these giant pulses have a common index and that they have different indices, need 
further consideration. In this connection, giant pulses detected from the 
millisecond pulsar B1821-24 \citep{romanijohnston2001}, despite the very limited 
quantity of data, may be significant. The data provide very weak evidence that 
the cumulative probability distribution (\ref{cdf_vela}) is power-law with 
$\beta \approx 3 - 5$ and so $\alpha = 7 - 11$. Although little confidence can be 
attached to these estimates prior to more extended observations, they hint that 
the range of $\alpha$ for giant pulses overlaps the values for giant micropulses and 
that the 
large inferred range for $\alpha$, from $\approx 4.5$ to at least $6.5$ is real. 
Since the three sets of observations for giant pulses 
are for widely different 
frequencies ($800$ versus $430$ and $1517$ MHz) it is possible that the apparent 
spread in $\alpha$ is real and corresponds to $\alpha$ increasing with 
observing frequency, reaching the range observed for giant micropulses 
only above 1 GHz. On the other hand, since the background of normal 
pulses is larger at lower frequencies, the apparent trend in 
$\alpha$ for giant pulses may be due to convolution of giant and 
``normal'' emission. 

   The foregoing observations and associated analyses imply, {\it prima facie}, that 
both giant pulses and giant micropulses can be interpreted in terms of power-law 
statistics with high values 
of $\alpha$ in the ranges $4.4 - 11$ and $6.5 \pm 0.7$, respectively. 
While the data suggest that $\alpha$ may increase with observing frequency 
for giant pulses but not giant micropulses (Table \ref{table_pulses}), these indices are similar enough to suggest a common 
interpretation for both giant phenomena. 

\section{Constraints on the Emission Mechanisms and Source Physics}

\subsection{Basic issues and SOC}

   Giant pulses and micropulses are frequently restricted to narrow phase windows 
compared with the normal pulse emission. They also occur as higher flux events 
than normal pulses and have observably different field statistics. These 
characteristics 
are strong evidence that giant pulses and micropulses are generated 
in a different source region and/or by a different emission process than
the normal pulsar emissions, which can almost certainly be interpreted 
in terms of SGT, based on analyses for three pulsars
\citep{cetal2001,cetal2002d,cetal2002a,cetal2002b}. One interpretation is that 
giant phenomena are generated in the neutron star's outer magnetosphere, 
associated with an outer gap plasma, while the normal pulses are 
generated in the inner magnetosphere near the inner gap 
\citep{romaniandyadigaroglu1995,romanijohnston2001}. 

   In general, the observed field statistics are determined by 
the intrinsic field statistics produced by relevant generation 
mechanisms, relativistic beaming effects, 
possible spatial variations across the source, and propagation effects. 
Relativistic beaming effects may be very important for the $P(E)$ distribution, due to intrinsic variations in 
Lorentz factor and emission direction inside the source (Section 2). 
However, until the field statistics are calculated for 
relativistic beaming, the importance of this effect cannot be 
evaluated quantitatively. Beaming effects are therefore neglected 
for the rest of this section. 

Consider next propagation effects and associated refractive 
lensing by density irregularities. Ordinary 
propagation effects are discounted, since they likely produce approximately 
Gaussian intensity statistics 
\citep{ratcliffe1956,salpeter1967,rickett1977} and 
appear relatively unimportant for normal pulses 
\citep{cetal2001,cetal2002a,cetal2002b,cetal2002c}. Following 
Salpeter (1967) and Walker (personal communication, 2003) 
as described in Section 2, refractive lensing events should produce 
$P(F) \propto F^{-3}$, whence (1) and (2) imply $P(E)\propto E^{-5}$. Comparing 
this prediction with Table \ref{table_pulses}, lensing appears viable 
for giant pulses from the Crab pulsar, but marginally inconsistent with 
giant pulses from B1937+214 and B1821-24 (although $\alpha$ is very 
poorly known for the latter) and significantly inconsistent for the 
three known sources of giant micropulses. 
Moreover, this mechanism predicts that $\alpha$ should be 
independent of frequency, plausibly inconsistent with 
Table \ref{table_pulses}'s data for giant pulses. 
It may be relevant, however, that under some conditions the superposition 
of lensing events onto the distribution of normal pulses, and the 
reduced range of $F$ observed in the present datasets, may 
increase $\alpha$, perhaps  
as found for giant micropulses. In summary, at present the available 
data appear 
inconsistent with lensing and scattering effects being important 
for both giant pulses and giant micropulses, but are not yet 
definitive. 

    With relativistic beaming and refractive effects discounted 
at this time, the simplest interpretation is 
adopted: the observed field statistics are intrinsic (perhaps involving more than one mechanism) and not 
significantly influenced by spatial variations or relativistic 
beaming effects in the source. Progress can 
then be made in restricting the emission mechanism and source 
physics responsible for giant micropulses and pulses. The approximately 
power-law functional form of the observed field distribution is inconsistent 
with thermal waves, uniform secular growth, and elementary burst systems 
(modified Rayleigh, uniform, and exponential statistics, respectively) and 
with waves subject to nonlinear decay processes. Driven thermal 
waves, while they develop power-law statistics at high $E$ \citep{r1995,cetal2000}, 
are a most unlikely explanation since the observed 
brightness temperatures are many orders of magnitude larger than plausible thermal 
temperatures. 
  
   SOC might appear attractive since it predicts power-law statistics. 
However, the power-law indices estimated directly are $\approx 4 - 7$, 
differing greatly from $1$ and lying outside the range $0.5 - 3$ typical of phenomena 
interpreted in terms of SOC \citep{bak1996}. Accordingly, it is doubtful that SOC 
applies to giant pulses and micropulses, contrary to an earlier qualitative suggestion 
\citep{youngkenny1996}. Reversing this conclusion requires that (1) the index for the 
distribution of giant micropulses/pulses is much lower than that estimated directly, 
perhaps due to the convolution effects mentioned in subsection 5.3 below, and/or 
(2) a theoretical model based 
on SOC can be developed with such large power-law indices. 

\subsection{Wave collapse}

   The nonlinear self-focusing process of wave collapse leads directly to 
power-law statistics at high $E$ \citep{rn1990,r1996} and so 
is immediately attractive for giant pulses and micropulses. Moreover, the 
indices $4.4 - 7.3$ (including the error bounds) 
determined directly for the four pulsars 
with reasonably well-determined field statistics (Table \ref{table_pulses} 
excluding PSR B1821-24) overlap with the range $4 - 7$ predicted for wave 
collapse in non-relativistic, electrostatic, 
electron-proton simulations \citep{rn1990,r1996}. In particular, 
substituting $D = 2$ or $3$ into the 
results in Table \ref{table_theory}, isotropic collapse theory predicts 
$\alpha = 4$ and $5$, with the corresponding predictions for 
prolate and oblate wavepacket shapes (appropriate to magnetized conditions)
being $\alpha = 5$, $6$ and $5$, $7$, respectively. 
Accordingly, interpretations in terms of modulational instability 
and wave collapse appear viable for giant micropulses and pulses: earlier work for 
normal pulses \citep{weatherall1997,weatherall1998} may thus be relevant to 
giant phenomena instead. Very recent analyses of short timescale 
structures observed in 
giant pulses from the Crab pulsar \citep{hankinsetal2003} 
provide qualitative support for the conclusion reached here. 

   Three points of concern should be raised, however. First, 
although waves undergoing collapse in 
relativistic, electron-positron, strongly magnetized plasmas 
appropriate to pulsar magnetospheres are expected to have 
power-law statistics, based on the strong qualitative 
similarities with collapse in electron-proton plasmas 
\citep{weatherall1997,weatherall1998}, the 
power-law indices are not yet known. Specifically, the very 
different wave dispersion, nonlinearities, magnetization effects, and 
electromagnetic character 
\citep{melrosegedalin1999,gedalinetal2002} could well alter the collapse 
dynamics and so the predicted indices from those known 
\citep{r1997} for non-relativistic electron-proton simulations. 
More work is thus required to make wave collapse into a fully 
quantitative theory for giant pulses and micropulses. 

Second, weak evidence exists (Section 4) that 
$\alpha$ increases with observing frequency for giant pulses but 
not giant micropulses, corresponding either to an increase 
in $D$ and/or a change in wavepacket shape. While this may be possible, the 
standard model of pulsar magnetospheres is that the plasma density and magnetic 
field strength decrease monotonically with distance from the pulsar (once 
well above the inner gap region), whence it is more plausible that $D$ should 
increase, and magnetization effects decrease (corresponding to more isotropic 
wavepackets), with increasing height and so decreasing frequency. This 
qualitative argument predicts $\alpha$ should decrease with increasing 
frequency, opposite to the apparent trend in $\alpha$ for giant 
pulses. Given the possible importance of convolution effects, 
discussed more in subsection 5.3, the trend in $\alpha$ may be affected 
by sensitivity issues and variations in the observed flux relative to the 
receiver noise with frequency. This should be addressed in future 
multi-frequency observations.   

   Third, suppose two or more processes combine to produce the observed field 
statistics, for instance because collapsing wavepackets generate  
radiation at the electron plasma frequency and/or its harmonics via 
a nonlinear process or an antenna mechanism 
\citep{freundpapadopoulos1980,hafizigoldman1981,akimotoetal1988}. [Note that 
linear mode conversion, or direct escape of the collapsing wavepacket from the 
source region before ``burnout'', 
should not alter the field statistics from those predicted by collapse theory.] Then, combining a nonlinear process whose rate is proportional to, 
say, $W^{2}$ (or $E^{4}$) 
with the wavepacket statistics given by Table \ref{table_theory}, the 
overall statistics should still be power-law but with a larger  
index $\alpha'$ (ideally with $\alpha' = \alpha + 4$)  
than predicted in Table \ref{table_theory}. Further theoretical work is 
required to assess this possibility, for collapse, SOC, and potentially 
other theories.

\subsection{Convolution effects and SGT}

    It is emphasized now that closely power-law statistics for 
$1 - 2$ decades in 
$E$ ($2 - 4$ decades in $F$) can result \citep{cetal2002d,cetal2002b} from 
vector superposition of a Gaussian (intensity) or lognormal distribution 
with a lognormal distribution centered at lower $E$ but extending to 
higher $E$ (due to larger $\sigma$ in equation \ref{p_sgt}). Exactly 
this effect appears relevant in the ``transition regions'' of three pulsars, explaining the statistics of normal pulses at pulsar phases 
where the normal pulsar emission is just becoming observable 
above the background \citep{cetal2002a,cetal2002b,cetal2002c}. 
Table \ref{table_pulses} shows that the giant 
pulses and micropulses are observed over at most 2 decades in $F$. Accordingly, 
two crucial observational questions remain to be answered: 
(1) are the apparently power-law statistics observed for these giant 
phenomena intrinsic or are they due to vector superposition of two or more 
distributions (with the combination of a 
lognormal with either a Gaussian intensity distribution or another lognormal known 
to produce the effect)? (2) If intrinsic, are the power-law indices significantly 
affected by superposition of the giant micropulses with the distribution of 
normal pulses and/or receiver noise? 

   These issues therefore affect whether SGT is a viable theory 
for giant pulses and micropulses.  A single wave population obeying 
pure SGT is predicted to yield lognormal statistics; this 
simple interpretation is then inconsistent with the data. However, 
there are at least two ways in which the apparently power-law 
statistics for giant pulses and micropulses can be reconciled 
with SGT. First, as pointed out just above, vector convolution 
of a lognormal with either a Gaussian intensity distribution 
(e.g., sky background and receiver noise) or a second lognormal 
can result in field statistics that appear closely power-law 
for a broad range of high fluxes/fields 
\citep{cetal2002d,cetal2002b,cetal2002c}. Second, SGT can 
coexist at moderate $E$ with wave collapse at 
high $E$ \citep{r1995,cetal2002a}, resulting in a power-law 
distribution at high 
enough $E$ (above the peak in the SGT distribution), as discussed in Section 2. 

    Longer duration observations which determine the field statistics over larger 
ranges of $F$ are required to answer these questions definitively. 
Reasons are that the extent of the range in $F$ over which the statistics are 
approximately power-law will become apparent, as will changes in the statistics at 
large $F$. The second reason relates to the fact that if the apparently 
power-law statistics are due to vectorial superposition of two non-power-law 
distributions or one power-law distribution with a non-power-law distribution, then 
at high enough $F$ the functional form of the 
combined distribution will evolve to that of the component  
distribution which dominates at large $F$ \citep{cetal2002d}.

\section{Summary and Conclusions}

(1) The different statistical properties of giant pulses and 
micropulses from the normal pulsar emission, and the restricted phase windows in which they are typically observed (except for the 
Crab pulsar and B0950+08), imply that the giant and normal emissions 
are most likely produced in distinct source regions via different 
emission mechanisms. 

(2) The power-law indices estimated directly from the field distributions for the three known sources of giant micropulses at $0.4$, $1.5$ and 
$2.3$ GHz are almost identical at $6.5 \pm 0.7$, so that 
they may be regarded as one population with similar source physics. 

(3) The power-law indices estimated directly from observations of known 
sources of giant pulses are $4.6 \pm 0.2$, $5.6 \pm 0.6$, and 
$7 - 11$ at 430, 800, and 1.5 GHz, respectively. The indices may 
increase with observing frequency, perhaps reaching the values for 
giant micropulses above 1 GHz. If real this trend suggests a difference from giant micropulses, but the trend may be due to the relative 
background varying with frequency. 

(4) The power-law indices of giant pulses and giant micropulses appear 
to be sufficiently similar for them to have a common theoretical 
interpretation. 

(5) It is not yet established conclusively that the apparently 
power-law field statistics of giant micropulses and pulses are 
intrinsic rather than the 
result of vectorially convolving two non-power-law distributions. Such 
non-intrinsic power-law statistics are found for normal pulses in the 
transition regions where the normal pulsar emissions 
are emerging above the background \citep{cetal2002b,cetal2002c}. 
Extended observations of giant micropulses and pulses are required, with 
fitting of at least two wave components, to determine whether 
the observed power laws are intrinsic at high $F$ 
and, if so, what the true power-law indices are. This requires longer-duration  
observations so as to obtain better statistics and observe events 
over a larger range of fluxes. 

(6) Relativistic beaming effects, which produce variations in 
observed flux due to changes in Lorentz factor and the direction 
of maximum emission relative to the observer (e.g., due to 
magnetic turbulence or changes in the plasma flow velocity), might 
be relevant. Theoretical models for the ensuing flux statistics 
are required, with high priority since pulsar plasmas are certainly 
relativistic and pulsar emissions are clearly beamed. 
  
(7)  Emission mechanisms involving thermal waves, driven thermal waves, 
uniform secular growth, elementary burst systems, and nonlinear decay 
mechanisms are inconsistent with the 
functional form of the observed distribution. The first two are also implausible 
on energy grounds. Lensing events due to refraction by density irregularities 
are interesting since they produce power-law field statistics (M.A. Walker, 
personal communication, 2003) with $\alpha = 5$. This 
value is consistent with giant pulses from only one pulsar (the Crab), but not 
the giant micropulses or the apparent variation of $\alpha$ with observing 
frequency. Observational limitations mean that this mechanism cannot 
yet be ruled out. 

(8) The power-law indices $\approx 4.4 - 7$ estimated directly from the 
observed distributions differ greatly from the value $\approx 1$ for 
simple SOC theory and lie outside the range $\approx 0.5 - 3$ of known 
SOC systems. SOC is therefore implausible despite the statistics being 
power-law. Reversing this conclusion requires, first, further 
observations confirming that the giant phenomena have intrinsic 
power-law statistics whose indices were significantly 
overestimated due to convolution effects with the normal pulses and 
measurement noise and/or, second, construction of theoretical SOC models with 
suitably large values of $\alpha$. 

(9) Nonlinear modulational instabilities and wave collapse processes typically 
have power-law field statistics, qualitatively consistent with the available 
data. Moreover, the observed indices $\alpha \approx 4.4 - 7$ 
lie within the range $4 - 7$ predicted by current theories for wave 
collapse \citep{rn1990,r1996}. This is currently the most 
plausible mechanism for giant pulses and micropulses, 
complementing another argument based on timescales 
\citep{hankinsetal2003}. 

(10) Further research on applying collapse theory and modulational 
instabilities to giant pulses and micropulses 
is required to resolve open issues relating to (i) whether 
the field statistics for collapse in 
the relativistic, electron-positron, highly 
magnetized plasmas in pulsar magnetospheres differ from those 
known for non-relativistic electron-proton plasmas, (ii) explaining the 
apparent trend in $\alpha$ with observing frequency for giant 
pulses being opposite to that predicted qualitatively 
based on magnetization 
effects and current collapse theory, and (iii) considering in 
detail how collapse produces the observed radiation and how 
any ancillary radiation mechanism modifies the predicted field 
statistics. 

(11) Finally, SGT cannot yet be ruled out as important in 
understanding giant pulses and micropulses since (i) the 
observed power-law statistics can be interpreted in terms of vector convolution 
of one or more of the lognormal distributions predicted for SGT, 
as already observed for normal pulses 
\citep{cetal2002d,cetal2002b,cetal2002c}, 
and (ii) wave collapse and SGT can coexist. The differences between 
normal and giant pulsar emission favor, but do not require, a different 
theoretical interpretation for the two phenomena, thereby favoring 
wave collapse over SGT for giant pulses and micropulses.   

    In conclusion, progress has been made in 
determining the intrinsic field statistics of giant micropulses and giant 
pulses and identifying therefrom the source physics and emission 
mechanisms. Both phenomena appear to have very similar field statistics and 
to admit a common theoretical interpretation, although whether the
observed power-law 
features are intrinsic or due to convolution effects remains to be 
determined. The nonlinear process of wave collapse produces power-law 
statistics with indices that can be in the observed range and, despite some 
theoretical difficulties, appears the most plausible theoretical 
interpretation. SGT remains viable although less favored, while SOC, 
refractive lensing, and certain other mechanisms appear not viable. 
Relativistic beaming effects due to changes in Lorentz factor and 
emission direction still need to be investigated and are plausibly 
very important. Longer duration, more sensitive 
observations of giant phenomena over a larger range of fluxes, together 
with associated fitting of multiple vectorially-convolved wave distributions 
should resolve these observational issues. Outstanding theoretical 
issues may be resolved by extending current simulations and 
theories for relativistic beaming and wave collapse, together with
 any other radiation mechanisms required, 
to conditions appropriate for pulsar magnetospheres.  


%% Included in this acknowledgments section are examples of the
%% AASTeX hypertext markup commands. Use \url without the optional [HREF]
%% argument when you want to print the url directly in the text. Otherwise,
%% use either \url or \anchor, with the HREF as the first argument and the
%% text to be printed in the second.

\acknowledgments

I thank the Referee for emphasizing the possible importance 
of relativistic beaming effects, M.~A. Walker for personal 
communications involving refractive lensing events, and 
S. Johnston, M.~A. Walker, and P.~A. Robinson 
(all at U. Sydney) for helpful discussions and comments on this 
paper. The Australian Research Council funded this research.




%\bibitem[King(1966)]{kin66}  King, I. R.  1966, \aj, 71, 276
%\bibitem[King(1975)]{kin75}  King, I. R.  1975, Dynamics of
%   Stellar Systems, A. Hayli, Dordrecht: Reidel, 1975, 99

\clearpage



\begin{thebibliography}{}
\bibitem[Akimoto et al.(1988)]{akimotoetal1988} Akimoto, K., Rowland, H.~L., 
\& Papadopoulos, K. 1988, Phys. Fluids, 31, 2185

\bibitem[Asseo(1996)]{asseo1996} Asseo, E. 1996, in ASP Conf Ser. 105, 
Pulsars: Problems and Progress, ed. S. Johnston, M. A. 
Walker, \& M. Bailes (San Francisco: ASP), 147

\bibitem[Asseo et al.(1990)]{asseoetal1990} Asseo, E., Pelletier, G., \& 
Sol, H. 1990, MNRAS, 247, 529 

\bibitem[Bak(1996)]{bak1996} Bak, P. 1996, How Nature Works, 
(Copernicus, New York).  

\bibitem[Bak et al.(1987)]{baketal1987}
Bak, P., Tang, C., \& Weisenfeld, K. 1987, \prl, 59, 381 

\bibitem[Bak et al.(1988)]{baketal1988}
Bak, P., Tang, C., \& Weisenfeld, K. 1988, Phys. Rev. A, 38, 364
 
\bibitem[Cairns \& Grubits(2001)]{cg2001} Cairns, I.~H., \& Grubits, K.~A. 2001, 
Phys. Rev. E, 64, 056408

\bibitem[Cairns \& Menietti(2001)]{cm2001} Cairns, I.~H., \& Menietti, J.~D. 2001, 
J. Geophys. Res., 106, 29515

\bibitem[Cairns \& Robinson(1999)]{cr1999} Cairns, I.~H., \& Robinson, P.~A. 1999, 
\prl, 82, 3066

\bibitem[Cairns et al.(2000)]{cetal2000} Cairns, I.~H., Robinson, P.~A., \&  
Anderson, R.~R. 2000, \grl, 27, 61 

\bibitem[Cairns et al.(2001)]{cetal2001} Cairns, I.~H., 
Johnston, S., \& Das, P. 2001, ApJL, 563, L65

\bibitem[Cairns et al.(2002)]{cetal2002d} Cairns, I.~H., Robinson, P.~A., 
\& Das, P. 2002, Phys. Rev. E, 66, 066614

\bibitem[Cairns et al.(2003a)]{cetal2002a} Cairns, I.~H., Das, P., 
Johnston, S., \& Robinson, P.~A., 2003a, MNRAS, 343, 512

\bibitem[Cairns et al.(2003b)]{cetal2002b} Cairns, I.~H., Das, P., 
Johnston, S., \& Robinson, P.~A., 2003b, MNRAS, 343, 523

\bibitem[Cairns et al.(2003c)]{cetal2002c} Cairns, I.~H., Johnston, S., 
\& Das, P.,2003c, MNRAS, submitted

\bibitem[Cognard et al.(1996)]{cognardetal1996} Cognard, I., Shrauner, J.~A., 
Taylor, J.~H., \& Thorsett, 
S.~E. 1996, \apjl, 457, L81

%\bibitem[Craft et al.(1968)]{craftetal1968}
% Craft, H.~D., Comella, J.~M., \& Drake, F.~D. 1968, Nature, 
%218, 1122 

%\bibitem[Drake \& Craft(1968)]{drakecraft1968}
%Drake, F.~D.,  and Craft, H.~D. 1968,  
%Nature,  220, 231 

\bibitem[Freund and Papadopoulos(1980)]{freundpapadopoulos1980}
Freund, H.~P., and Papadopoulos, K.~D. 1980, Phys. Fluids, 23, 732

\bibitem[Gedalin et al.(2002)]{gedalinetal2002} Gedalin, M.E., 
Gruman, E., \& Melrose, D.B. 2002, MNRAS, 337, 422

\bibitem[Hafizi \& Goldman(1981)]{hafizigoldman1981}
Hafizi, B., \& Goldman, M.~V. 1981, Phys. Fluids, 24, 145
 
\bibitem[Hankins(1996)]{hankins1996} Hankins, T.~H. 1996, in ASP Conf. Ser. 105, 
Pulsars: Problems and Progress, ed. S. Johnston, M.~A. Walker, \& M. Bailes (San 
Francisco: ASP), 197 

\bibitem[Hankins et al.(2003)]{hankinsetal2003} Hankins, T.~H., Kern, J.~S., 
Weatherall, J.~C., \& Eilek, J.~A. 2003, Nature, 422, 141 

\bibitem[Heiles et al.(1970)]{heilesetal1970}
Heiles, C., Campbell, D.B., \& Rankin, J.M., Nature, 226, 529 

%\bibitem[Isliker \& Benz(2001)]{islikerbenz2001}
%Isliker, H., and Benz, A. 2001, A.\&A., 375, 1040  

\bibitem[Johnston \& Romani(2002)]{johnstonromani2002}
Johnston, S., \& Romani, R. 2002, \mnras, 332, 109 

\bibitem[Johnston et al.(2001)]{johnstonetal2001}
Johnston, S.,  van Straten, W., Kramer, M., \& Bailes, M. 2001,  
\apjl, 549, L101 

\bibitem[Kramer et al.(2002)]{krameretal2002} Kramer, M.,
van Straten, W., Johnston, S., \& Bailes M., 2002, \mnras, 334, 523

%\bibitem[Kuznetsov et al.(1986)]{kuznetsovetal1986} Kuznetsov, E.A., Rubenchik, A.M., 
%\& Zakharov, V.E., 1986, Phys. Rep., 142, 103 

\bibitem[Lundgren et al.(1995)]{lundgrenetal1995} Lundgren, S.C., 
Cordes, J.M., Ulmer, M., Matz, S.M., Lomatch, S., Foster, R.S., \& Hankins, T.,  1995, 
ApJ, 453, 433
 
\bibitem[Luo \& Melrose(1995)]{luomelrose1995} Luo, Q., \& Melrose, D.B. 1995, 
\mnras, 276, 372 

\bibitem[Manchester \& Taylor(1977)]{man77} Manchester, R.~N., \& Taylor, J.~H. 1977,  
Pulsars (San Francisco:Freeman), 281
 
\bibitem[Melrose(1996)]{melrose1996} Melrose, D.~B. 1996, in ASP Conf. Proc. 105, 
Pulsars: Problems and Progress, ed. S. Johnston, M. A. 
Walker, \& M. Bailes et al. (San Francisco: ASP), 139

\bibitem[Melrose \& Gedalin(1999)]{melrosegedalin1999} Melrose, D.B., \&
Gedalin, M.~E. 1999, ApJ, 521, 351 

\bibitem[Pelletier et al.(1988)]{pelletieretal1988} Pelletier, G., Sol., H., \& 
Asseo, E. 1988, Phys. Rev. A, 38, 2552

\bibitem[Press et al.(1986)]{pressetal1986} Press, W.~H., Flannery, B.~P., 
Teukolsky, S.~A., \& 
Vetterling, W.~T. 1986, Numerical 
Recipes (New York, Cambridge) 

\bibitem[Queinnec \& Zarka(2001)]{queinneczarka2001}
Queinnec J., \& Zarka, P. 2001, Plan. Space Sci., 49, 365 

\bibitem[Ratcliffe(1956)]{ratcliffe1956} Ratcliffe, J.~A. 1956, 
Rep. Prog. Phys., 19, 188

\bibitem[Rickett(1977)]{rickett1977} Rickett, B.~J., 1977 
Ann. Rev. Astron. \& Astrophys., 15, 471

%\bibitem[Rickett(1990)]{rickett1990} Rickett, B.~J., 1990 
%Ann. Rev. Astron. \& Astrophys., 28, 561 

\bibitem[Robinson(1992)]{r1992}
Robinson, P.A. 1992, Sol. Phys., 139, 147

\bibitem[Robinson(1995)]{r1995}
Robinson, P.A. 1995, Phys. Plasmas, 2, 1466

\bibitem[Robinson(1996)]{r1996}
Robinson, P.A. 1996, Phys. Plasmas, 3, 192

\bibitem[Robinson(1997)]{r1997} Robinson, P.A. 1997, Rev. Mod. Phys., 69, 507

\bibitem[Robinson \& Cairns(2001)]{rc2001}
Robinson, P.~A., \& Cairns, I.~H. 2001, Phys. Plasmas,  8, 2394

\bibitem[Robinson \& Newman(1990)]{rn1990} Robinson, P.A., \&
Newman, D.L. 1990, Phys. Fluids B, 2, 2999

\bibitem[Robinson et al.(1993)]{retal1993}
Robinson, P.~A., Cairns, I.~H., \& Gurnett, D.~A 1993, \apj, 407, 
790 

\bibitem[Robinson et al.(1996)]{rsw1996}
Robinson, P.~A., Smith, H.~B., and Winglee, R.~M. 1996, \prl, 76,  
3558 

\bibitem[Romani and Johnston(2001)]{romanijohnston2001} 
Romani, R., \& Johnston, S. 2001, \apj, 557, L93

\bibitem[Romani \& Yadigaroglu(1995)]{romaniandyadigaroglu1995}
Romani, R., \& Yadigaroglu, I.-A. 1995, \apj, 438, 314

\bibitem[Sallmen et al.(1999)]{sallmenetal1999} 
Salmen, S., Backer, D.~C., Hankins, T.~H., Moffett, D., and 
Lundgren, S. 1999, \apj, 517, 460 

\bibitem[Salpeter(1967)]{salpeter1967} 
Salpeter, E.E. 1967, ApJ, 147, 433 

\bibitem[Schneider et al.(1992)]{schneideretal1992}
Schneider, P., Ehlers, J., \& Falco, E.~E. 1992, Gravitational Lenses, 
(Springer-Verlag, Berlin) 

\bibitem[Staelin \& Reifenstein(1968)]{staelinreifenstein1968}
Staelin, D.~H., \& Reifenstein, E.~C. 1968, Science, 162, 1481

\bibitem[Weatherall(1997)]{weatherall1997}
Weatherall, J.~C. 1997, \apj, 483, 402

\bibitem[Weatherall(1998)]{weatherall1998}
Weatherall, J.~C. 1998, \apj, 506, 341

\bibitem[Young \& Kenny(1996)]{youngkenny1996} Young, M.~D.~T., and Kenny, B.~G. 1996, 
in ASP Conf Ser.  105, Pulsars: Problems and Progress, ed. S. Johnston, M.A. 
Walker, \& M. Bailes (San Francisco, ASP), 179 
\end{thebibliography}
\end{document}